\documentclass{comnet}

\usepackage{amsmath} 
\usepackage{amsthm} 
\usepackage{graphicx} 
\usepackage[utf8]{inputenc}

\graphicspath{{pics/}}
\usepackage[usenames, dvipsnames]{color}
\usepackage[]{algorithmic}
\usepackage[]{algorithm}
\usepackage{mathtools}

\def\vec#1{\ensuremath{\mathchoice
                     {\mbox{\boldmath$\displaystyle\mathbf{#1}$}}
                     {\mbox{\boldmath$\textstyle\mathbf{#1}$}}
                     {\mbox{\boldmath$\scriptstyle\mathbf{#1}$}}
                     {\mbox{\boldmath$\scriptscriptstyle\mathbf{#1}$}}}}%

\begin{document}

\title{
Geometric Multiscale Community Detection: Markov Stability and Vector Partitioning
}

\shorttitle{Multiscale Community Detection and Vector Partitioning} 
\shortauthorlist{Liu and Barahona} 

\author{
\name{Zijing Liu$^*$}
\address{Departments of Mathematics and Chemistry, Imperial College London, South Kensington Campus, London SW7 2AZ, UK\email{$^*$ zijing.liu@imperial.ac.uk}}
\and
\name{Mauricio Barahona$^{\dagger}$}
\address{Department of Mathematics, Imperial College London, South Kensington Campus, London SW7
2AZ, UK\email{$^\dagger$ m.barahona@imperial.ac.uk}}
}

\maketitle

\begin{abstract}
{Multiscale community detection can be viewed from a dynamical perspective within the Markov Stability framework, which uses the diffusion of a Markov process on the graph to uncover intrinsic network substructures across all scales. Here we reformulate multiscale community detection as a max-sum length vector partitioning problem with respect to the set of time-dependent node vectors expressed in terms of eigenvectors of the transition matrix.
This formulation provides a geometric interpretation of Markov Stability in terms of a time-dependent spectral embedding, where the Markov time acts as an inhomogeneous geometric resolution factor that zooms the components of the node vectors at different rates. 
Our geometric formulation encompasses both modularity and the multi-resolution Potts model, which are shown to correspond to vector partitioning in a pseudo-Euclidean space, and is also linked to spectral partitioning methods, where the number of eigenvectors used corresponds to the dimensionality of the underlying embedding vector space.
Inspired by the Louvain optimisation for community detection, we then propose an algorithm based on a graph-theoretical heuristic for the vector partitioning problem. We apply the algorithm to the spectral optimisation of modularity and Markov Stability community detection.
The spectral embedding based on the transition matrix eigenvectors leads to improved partitions with higher information content and higher modularity than the eigen-decomposition of the modularity matrix. We illustrate the results with random network benchmarks.}
{multiscale community detection, spectral methods, partitioning algorithms, modularity, Markov stability.}
\\
\end{abstract}

\section{Introduction}
\label{sec:1}

Networks provide a concise way to represent relational and structural information of models and data, 
and to link function with structure. 
Community detection can help reveal the relationships inherent in complex networks by finding groups of nodes in the graph that are strongly related within the group, and much less so across groups. 
From this perspective, vertices (or nodes) in the same community are `close' in a structural sense, signifying that the objects represented by the vertices in each community share similar function or qualities. 
Community detection in complex networks has attracted much attention due to its potential for practical applications, as well as its close mathematical relation to problems such as data clustering, graph partitioning, and image segmentation~\cite{girvan2002community,porter2009communities,fortunato2010community,delvenne2013stability}.

Many community detection methods have been proposed in the past few years. A group of widely used methods is based on the optimisation of modularity~\cite{newman2004finding}, a well-known quality measure for community detection. Modularity maximisation can also be seen as a particular case of maximum likelihood of the degree-corrected stochastic block model~\cite{newman2016equivalence}. In its original formulation, modularity was restricted to finding a unique community partition, at a particular scale. However, complex networks commonly have modular structures at several levels of resolution, and a `one-shot community detection approach' is not appropriate to describe the network connectivity. Furthermore, modularity is limited both by a resolution limit~\cite{fortunato2007resolution} and a field-of-view limit~\cite{schaub2012markov}, which preclude the detection of communities smaller or larger than those limits, respectively, or with  sparse and non-clique-like community structure~\cite{schaub2012markov}. 

One way to alleviate these issues is to take a `full-screening community detection approach' by zooming across scales to detect intrinsic structure in the network. Markov Stability provides such a framework by using the inhomogeneity of the diffusion of a Markov process on the network across time as a means to scan the graph and reveal community structure at different resolution levels~\cite{delvenne2010stability,lambiotte2014random}. Interestingly, modularity can be recovered as a special case of Markov Stability (corresponding to a one-step, discrete-time random walk), whereas the (normalised) Fiedler bipartitioning is obtained as the diffusion time (i.e., the number of steps) goes to infinity~\cite{delvenne2013stability}. As a generalised framework, Markov Stability has been applied to a variety of problems, including protein structure~\cite{delmotte2011protein,amor2014uncovering}, social networks~\cite{beguerisse2014interest} and neuronal network analyses~\cite{bacik2016flow}, among others~\cite{lambiotte2014random}.

As is generally the case for graph partitioning algorithms, the optimisation of Markov Stability is NP-hard, and only approximate optimisation methods can be used computationally for large networks. An important class of such optimisation approaches are spectral methods~\cite{newman2013spectral}, which are based on the eigen-decomposition of a matrix related to the graph, e.g., the adjacency matrix, the Laplacian matrix or the modularity matrix~\cite{newman2006modularity}. Indeed, spectral methods based on the adjacency matrix can be treated as a relaxation of the graph min-cut problem~\cite{kernighan1970efficient}, which is a well-known way to formulate the graph partition problem~\cite{von2007tutorial}. 
Similarly, the normalised graph cut problem can be relaxed in a continuous domain by the eigen-decomposition of the normalised Laplacian matrix~\cite{shi2000normalized}, whereas spectral methods of the combinatorial graph Laplacian matrix are related to the ratio-cut problem~\cite{hagen1992new}. 
The theoretical connections with spectral methods are fundamentally appealing, yet when it comes to the optimisation of modularity (or Markov Stability), such relationships are less clear and several problems arise. One principal issue is that spectral methods are single-scale methods and need the number of communities as a prior declaration. Therefore, previously used spectral algorithms for modularity maximisation find the number of communities either through a divisive bipartition scheme or through exhaustive search~\cite{newman2006modularity,white2005spectral}. 

In this work, we show that Markov Stability optimisation is equivalent to a max-sum vector partitioning problem of the embedding of the nodes of the graph in a geometric vector space~\cite{alpert1995spectral,onn2001vector}. This formulation provides a geometric interpretation for community detection, linking spectral methods and Markov Stability optimisation, and elucidates the role of time of the Markov process as a geometric resolution scale. Our reformulation also allows us to propose an optimisation heuristic for the vector partitioning problem inspired by the graph-theoretical node agglomeration notions used in the efficient, widely used Louvain optimisation method~\cite{blondel2008fast}. This implementation returns the optimised number of communities as an \emph{output} of the algorithm. 
We also show that our formulation encompasses both modularity and the Reichardt \& Bornholdt Potts model~\cite{reichardt2006statistical} for community detection. Compared to using the eigenvectors of the modularity matrix, our formulation achieves improved modularity with fewer eigenvectors, and with communities that have a better quality (lower uncertainty) when compared to the ground truth in random benchmarks.

\section{A Geometric Reformulation of Markov Stability through Vector Partitioning}
\label{sec:2}

The vector partitioning problem is defined as the partitioning of a set of $n$ vectors into $c$ groups such that an objective function of the sum of vectors in each group is maximised. There is a close connection between vector partitioning and graph partitioning. For example, the minimum cut problem of a graph can be formulated as a vector partitioning problem using the eigenvectors of the graph Laplacian~\cite{alpert1995spectral,alpert1999spectral}. More recently, it has been shown that the optimisation of modularity can be approximated by a vector partitioning problem via the eigen-decomposition of the modularity matrix~\cite{zhang2015multiway}. 
However, this approximation to modularity optimisation does not apply naturally to the analysis of Markov Stability, which relies on time-dependence for multiscale community detection. 
More specifically, the spectral decomposition should be compatible with the time evolution so that the eigenvalue problem is only solved once.

Below we introduce a set of vectors in $\mathbb{R}^{n-1}$, which are parameterised by the Markov time, and show that Markov Stability optimisation is equivalent to a max-sum partitioning of these vectors. This embedding provides a geometric interpretation of the temporal scale in Markov Stability: as time progresses the vectors in the basis shrink at different rates leading to an inhomogeneous resolution scale for multi-scale community detection.

\subsection{Markov Stability and vector partitioning}

Let us consider an undirected, weighted, connected graph of $n$ nodes (or vertices) with the $n \times n$ adjacency matrix $A$, where $A_{ij}=A_{ji}>0$ is the weight of the edge connecting vertices $i$ and $j$ and $A_{ij}=A_{ji}=0$ if there is no edge connecting vertices $i$ and $j$. The degree of the vertices is compiled as an $n$-dimensional vector $\vec{d}$ with components $d_i = \sum_{j=1}^n A_{ij}$. We also define the degree matrix $D$, an $n \times n$ diagonal matrix with the degrees of the vertices on its diagonal ($D_{ii} = d_i$). The total weight of the degrees of the networks is $m = \sum_{i,j} A_{ij}/2$.

\paragraph{Markov Stability \newline} Given the matrix $A$, we define a continuous-time Markov process taking place on the network and governed by the dynamics:
\begin{equation}
\frac{d\vec{p}}{dt} = -\vec{p} \, (I-D^{-1}A) := -\vec{p} \, (I-M),
\label{eq:Markov_proc}
\end{equation}
where $\vec{p}$ is an $1 \times n$ row vector defined on the vertices, $I$ is the identity matrix of size $n$, and $M$ 
is the transition matrix of the process. Note that $(I-M)$ is the random-walk normalised Laplacian matrix; hence~\eqref{eq:Markov_proc} is a diffusion process with a unique stationary distribution $\vec{\pi} = \vec{d}^T/2m$. 

The autocovariance matrix of $\vec{p}$ evolving under~\eqref{eq:Markov_proc} is given by 
\begin{align}
\label{eq:autocov}
B(t) = \Pi P(t) - \vec{\pi}^T\vec{\pi},
\end{align}
where $\Pi = D/2m $ and $P(t) = \mathrm{exp}(-t(I-M))$. The time $t$ of the process is denoted henceforth as the \emph{Markov time}~\cite{delvenne2013stability}.

Given a partition $g$ of the vertices into $c$ non-overlapping groups (or communities) denoted by $g=\{g_1,g_2,...,g_c\}$, the \textit{Markov Stability} of the partition is defined as~\cite{delvenne2010stability}:
\begin{equation}
\label{eq:MS}
 r(t,g) = \sum_{s=1}^{c} \, \sum_{i,j \in g_s} B(t)_{ij}. 
\end{equation}
Previous work has shown that by maximising Markov Stability, one can find optimised, robust partitions that are relevant over extended Markov times, and the sequence of optimised partitions across times so obtained reveals the multiscale community structure of the graph~\cite{delvenne2010stability, delmotte2011protein, delvenne2013stability, schaub2012markov, lambiotte2014random, beguerisse2014interest, bacik2016flow}.

\paragraph{Markov Stability as a geometric vector partitioning problem \newline}

We start by stating a proposition that follows directly from the definitions above, which allows for a straightforward spectral decomposition of the autocovariance matrix.

\begin{proposition}[Spectral relation between $B(t)$ and $M$]
\label{proposition}
Let us denote the eigenvalues of $M$ 
as $\lambda_1 = 1 \geq \lambda_2\geq \ldots \geq \lambda_n$ with corresponding eigenvectors $\vec{v}_k=[v_{k,1},v_{k,2}, \ldots ,v_{k,n}]^T$, such that $M\vec{v}_k=\lambda_k\vec{v}_k$ and $\vec{v}^{T}_k \Pi \vec{v}_l =\delta_{kl}$ where $\delta_{kl}$ is the Kronecker delta. 
Then the generalised eigenvalue problem 
\begin{align}
\label{eq:general_eigenvalue}
B(t)\vec{v} = \lambda \Pi \vec{v}
\end{align}
is solved by the eigenvectors 
$\{\vec{1},\vec{v}_2,...,\vec{v}_n\}$ with corresponding eigenvalues $\{0,\lambda_2(t),...,\lambda_n(t)\}$ with 
\begin{align}
\lambda_k(t) = \mathrm{exp}(-t(1-\lambda_k)) >0.
\label{eq:eigenvals}
\end{align}
\end{proposition}

The above proposition means that the autocovariance matrix $B(t)$ can be written as a Gram matrix:
\begin{equation}
B(t)_{ij} = \lambda_2(t) \pi_i v_{2,i} \pi_j v_{2,j} +  \lambda_3(t) \pi_i v_{3,i} \pi_j v_{3,j} + ... + \lambda_n(t) \pi_i v_{n,i} \pi_j v_{n,j}  := \vec{x}_i^T \vec{x}_j,
\label{eq:autocov_Gram}
\end{equation}
where we have defined the set of $(n-1)$-dimensional vectors $\vec{x}_i (t)$ which depend parametrically on the Markov time $t$:
\begin{align}
\vec{x}_i (t) = [\sqrt{\lambda_2(t)} \pi_i v_{2,i},\sqrt{\lambda_3(t)} \pi_i v_{3,i}, ...,\sqrt{\lambda_n(t)} \pi_i v_{n,i}]^T, \, \quad i=1, \ldots ,n.
\label{eq:vector}
\end{align}
In this formulation, each vertex of the network is thus represented by a time-varying `node vector' $\vec{x}_i (t)$ in a $(n-1)$-dimensional space. 

The Markov Stability~\eqref{eq:MS} of partition $g$ can then be rewritten as: 
\begin{equation}
\label{eq:VP}
r(t,g) 
 = \sum_{s=1}^c \, \sum_{i,j\in g_s} \vec{x}_i(t)^T \vec{x}_j (t) 
 = \sum_{s=1}^c \left \lVert{\sum_{i\in g_s} \vec{x}_i (t)} \right \rVert^2.
\end{equation}
Geometrically, this is equivalent to summing the `node vectors' within each community to generate a `community vector', and then computing the sum of the squared lengths of the community vectors. 

From this geometric rewriting, it follows that finding the partition that maximises Markov Stability is equivalent to finding the partition of the set of $n$ vectors $\vec{x}_i(t) \in \mathbb{R}^{n-1}$ that maximises the overall sum of the squared lengths of the sum vectors of each community. This is a \emph{max-sum length vector partitioning problem} with respect to the vectors $\vec{x}_i(t)$ given by~\eqref{eq:vector}.
This rewriting of Markov Stability allows us to investigate community detection as a geometric problem in the associated spectral vector space.

\paragraph{Markov Time as an inhomogeneous geometric resolution factor \newline}

As $t$ grows and the Markov process evolves on the network, all node vectors $\vec{x}_i(t)$ in the $(n-1)$-dimensional space approach the origin. 
This follows from the definition~\eqref{eq:vector}:
the $k$-th component of each vector is weighted by $\sqrt{\mathrm{exp}(-t(1-\lambda_{k+1}))}$ with $\lambda_{k+1} <1$ for $k=1, \ldots ,n-1$; hence all the components of the vector $\vec{x}_i(t)$ decay exponentially to zero as $t$ increases. 

In this setting, the Markov time $t$ acts as a geometric resolution factor that shrinks the node vectors as it increases.
However, the shrinking of the vectors is not homogeneous: different vector components decay at different rates, as determined by their associated eigenvalue
%
with smaller eigenvalues decaying faster. 
This inhomogeneity in the decay of each vector component induces changes in the \textit{relative geometry} of the set of eigenvectors as a function of Markov time, and hence to different optimal partitions at different Markov times. The different partitions obtained reveal the multi-scale community structure present in the graph, as illustrated with two simple constructive networks.

\begin{figure}[!h]
\centering
\includegraphics[width= 0.7\linewidth]{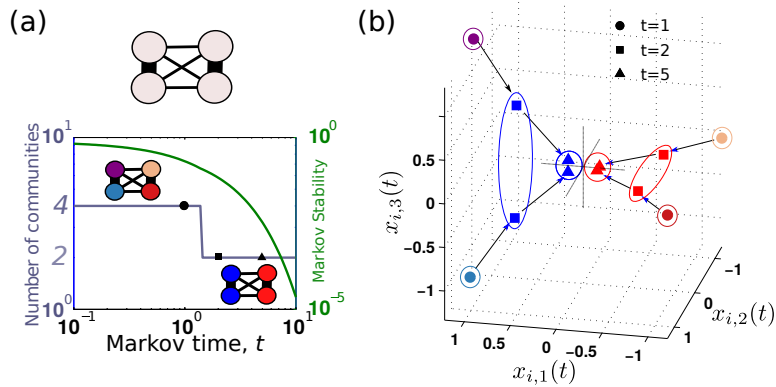}
\caption{(a) A simple network with $n=4$ vertices and one level of hierarchical structure:
the thicker edges have weight 10, and all other edges have weight 1. 
The sweeping of Markov time reveals the presence of a strong bipartition.  
(b) Visualisation of the four node vectors at Markov times $t=1$, $t=2$ and $t=5$. 
As time grows, the second and third components of $\vec{x}_i$ shrink faster than the first component, which leads to the emergence of the 2-way partition as time becomes larger when the angle between vectors becomes smaller than $\pi/2$ in this $(n-1)$-dimensional embedding.}
\label{fig:1}
\end{figure}

Figure~\ref{fig:1} shows the analysis of a simple network with four vertices and one level of hierarchical structure: two edges of the network have weight 10 and the other edges have weight 1.  
This simple example with $n=4$ allows us to visualise the full, unprojected $(n-1)$-dimensional node vectors (Fig.~\ref{fig:1}B). 
For small Markov times, every vertex forms its own community and the optimal number of communities is 4, but as
$t$ increases, an optimal partition into two communities is found. Geometrically, this can be understood by inspecting the node vectors.  At $t=1$, the angles between any two node vectors are all larger than $90^{\circ}$; hence grouping any two vectors does not increase the squared length of the vectors~\eqref{eq:VP} and each vector stays in its own group. As the Markov time increases (e.g., $t=2$ and $t=5$), the node vectors approach the origin but do so with inhomogeneous rates for the different vector components. This translates into angles less than $90^{\circ}$ between the node vectors corresponding to the nodes connected by the edge with large weight, thus revealing two communities in the network.

\begin{figure}[!h]
  \centering
  \includegraphics[width=0.7\textwidth]{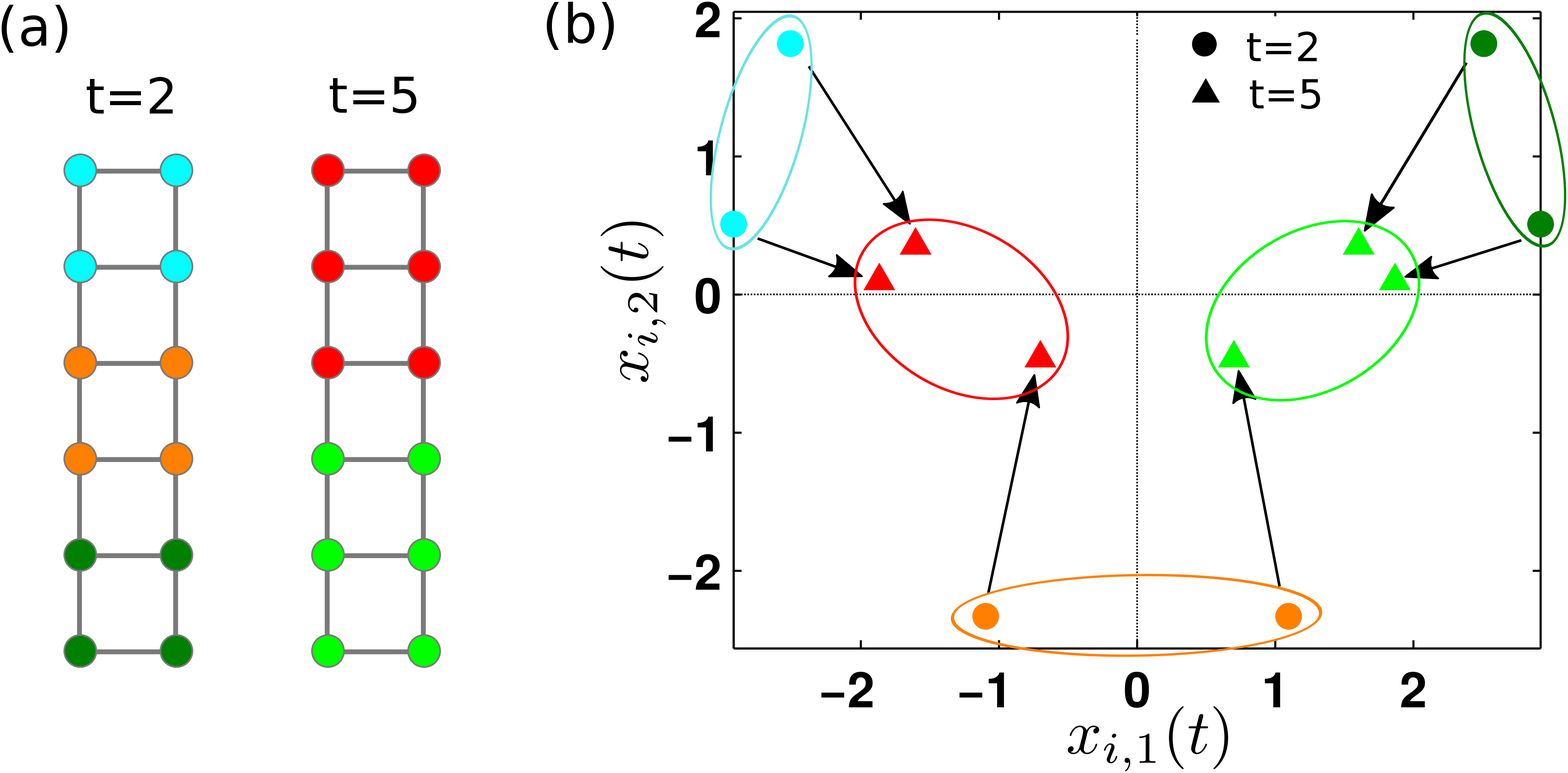}
  \caption{(a) A network without a strict hierarchy, as proposed in~\cite{reichardt2006statistical}. The network with nodes coloured according to the communities obtained at Markov times $t=2$ and $t=5$; (b) Visualisation of the node vectors (using the first two components) at Markov time $t=2$ and $t=5$. For this network, the optimum of the max-sum length vector partitioning problem changes from a 3-way partition to a 2-way partition as Markov time increases and the vectors $\vec{x}_i$ become closer to the origin and the angle between the node vectors in the orange community increase above  $90^{\circ}$.
Note that due to symmetry, only six (out of the twelve) node vectors $\vec{x}_i$ are visible, as some of the vectors overlap in this projection.}
  \label{fig:2}
\end{figure}

Our second example is a non-hierarchical network considered in Ref.~\cite{reichardt2006statistical}. 
Figure~\ref{fig:2} shows that in this case, a 3-way partition is obtained at Markov time $t=2$, whereas a 2-way partition is obtained at $t=5$. As $t$ increases, the angle between the vectors representing the vertices of the middle community (orange) increases above $90^{\circ}$ leading to the break up of this community into a 2-community structure.

As shown in~\cite{delvenne2013stability, delvenne2010stability}, the optimal partition for Markov Stability at large $t$ is given by the normalised Fiedler vector. Geometrically, this can be understood from
the leading 
asymptotics:
\begin{equation*}
\lim_{t\rightarrow +\infty} \vec{x}_i (t) =\sqrt{\lambda_2(t)} [ \pi_i v_{2,i},0, ...,0]^T \quad i=1, \ldots ,n.
\end{equation*}
i.e., the node vectors are dominated by their first component as $t \to \infty$. Since the optimal communities are obtained by grouping all vectors with angles between them below $90^{\circ}$, in this limit this corresponds to the bipartition according to the \textit{sign} of the entries of the second eigenvector $\vec{v}_2$---the normalised Fiedler vector associated with the second smallest eigenvalue of the normalised Laplacian.

\paragraph{What does $k$-means optimise in spectral clustering?\newline}
The $k$-means method is one of the most popular clustering methods applied on the usual spectral embedding. Yet its relationship with community detection is unclear in terms of their optimisation objectives.  Our vector representation allows us to establish the relationship between both approaches. In terms of $\vec{x}_i(t)$, the objective function that $k$-means minimises is:
\begin{equation}
\label{eq:kmeans}
\sum_{s=1}^c  \, \sum_{i\in g_s} \left \lVert { \vec{x}_i (t) - \bar{\vec{x}}_s} \right \rVert^2
 = \sum_{s=1}^c  \sum_{i\in g_s} \left \lVert { \vec{x}_i (t) } \right \rVert^2 -  \sum_{s=1}^c \frac{1}{\left \lvert {g_s} \right \rvert} \left \lVert {\sum_{i\in g_s} \vec{x}_i (t)} \right \rVert^2,
\end{equation}
where $\bar{\vec{x}}_s$ is the vector of the centroid of the group $g_s$ and $\left \lvert {g_s} \right \rvert$ is the size of the group $g_s$.
Note that the first term is a constant for a set of vectors at a given Markov time $t$, hence the objective of $k$-means is equivalent to maximising:
\begin{equation}
F(t,g)
 = \sum_{s=1}^c \frac{1}{\left \lvert {g_s} \right \rvert} \left \lVert {\sum_{i\in g_s} \vec{x}_i (t)} \right \rVert^2.
\end{equation}

Comparing to~\eqref{eq:VP}, it is easy to see that $k$-means maximises a \textit{normalised} version of the max-sum vector partitioning problem, where the squared length of the sum vector of a group is normalised by the group size.  Since modularity is a special case of Markov Stability, this explains why when the communities have equal sizes, k-means and vector partitioning have comparable performance for modularity. If, on the other hand, the community sizes are highly unequal, vector partitioning performs much better~\cite{zhang2015multiway}. Importantly, when the $k$-means objective~\eqref{eq:kmeans} is used, the number of communities need to be prescribed; otherwise, each vector will be partitioned into its own community since this will give an objective~\eqref{eq:kmeans} equal to zero.




\subsection{Geometric reformulation of modularity and Potts model for community detection}
\label{sec:4}
Previous work~\cite{delvenne2010stability, delvenne2013stability} has shown that community detection based on the popular single-resolution modularity~\cite{newman2004finding, newman2006modularity} and the related multi-resolution Potts model proposed by Reichardt \& Bornholdt~\cite{reichardt2006statistical} correspond to particular limits of Markov Stability. Specifically, minimising the Potts Hamiltonian for a partition $g$ is equivalent to maximising the \textit{linearisation} of Markov Stability:
\begin{align}
\label{eq:linearms}
r_{lin}(t,g) &= \sum_{s=1}^{c} \, \sum_{i,j \in g_s} B_{lin}(t)_{ij} 
\\  
\text{where} \quad \quad
 B_{lin}(t) &= \Pi \left[ (1-t)I + t M\right]  - \pi^T\pi.
\end{align}
In this linearisation, the Markov time $t$ is equivalent to the resolution $1/ \gamma$ in the Potts model~\cite{reichardt2006statistical}. When $\gamma=t=1$, the Potts model and linearised Markov stability are equivalent to modularity:
\begin{equation}
\mathrm{Modularity} = \sum_{s=1}^{c} \, \sum_{i,j \in g_s} B_{lin}(1)_{ij} 
\end{equation}
%

%

Using our geometric formulation, the Potts model (and modularity) can be recast as a vector partitioning problem in terms of the eigenvectors of the transition matrix. 
Similarly to PROPOSITION~\ref{proposition}, it is easy to show that the generalised eigenvalue problem of the linearised problem 
\begin{align}
B_{lin}(t)\vec{v} = \lambda \Pi \vec{v}
\end{align} 
is solved by the eigenvectors $\{\vec{1},\vec{v}_2,...,\vec{v}_n\}$ with corresponding eigenvalues $\{0,\mu_2(t),...,\mu_n(t)\}$ given by
\begin{align}
\mu_k(t) = 1-t (1- \lambda_k),
\end{align}
where the $\lambda_k$ are the eigenvalues of $M$.
Note that the eigenvalues $\mu_k(t)$ are no longer bounded and can become negative at different values of $t$.

For each Markov time, there will be a number $m$ $(1 \leq m \leq n)$ such that  $\mu_k(t)\geq0$ for  $1 \leq k \leq m$ and $\mu_k(t) < 0$ for $m < k \leq n$. We define this set of vectors in a pseudo-Euclidean space~\cite{shafarevich2012linear} with index of inertia $(m-1)$:
\begin{equation*}
\vec{x}_i (t) := [\sqrt{\mu_2(t)} \pi_i v_{2,i},\ldots,\sqrt{\mu_m(t)} \pi_i v_{m,i},\sqrt{-\mu_{m+1}(t)} \pi_i v_{m+1,i},\ldots,\sqrt{-\mu_{n}(t)} \pi_i v_{n,i}]^T \quad  i=1, \ldots ,n.
\end{equation*}
In this pseudo-Euclidean space, the magnitude of vector $\vec{x}_i (t)$ is given by:
\begin{equation*}
q\left(  \vec{x}_i (t) \right)  = \mu_2(t)\pi_i^2 v_{2,i}^2 + \ldots + \mu_m(t)\pi_i^2 v_{m,i}^2 + \mu_{m+1}(t)\pi_i^2 v_{m+1,i}^2 +  \ldots + \mu_n(t)\pi_i^2 v_{n,i}^2.
\end{equation*}

This allows us to write the following decomposition for $B_{lin}(t)$:
\begin{align*}
B_{lin}(t)_{ij} &= \mu_2(t) \pi_i v_{2,i} \pi_j v_{2,j} +  \mu_3(t) \pi_i v_{3,i} \pi_j v_{3,j} + ... + \mu_n(t) \pi_i v_{n,i} \pi_j v_{n,j}  \\ 
& = \frac{q(\vec{x}_i(t)+ \vec{x}_j(t)) - q(\vec{x}_i (t)) - q(\vec{x}_j (t))}{2}.
\end{align*}
The linearised stability~\eqref{eq:linearms} can then be written in terms of $\vec{x}_i (t)$ as:
\begin{equation}
\label{eq:PottsVP}
r_{lin}(t,g)  = \sum_{s=1}^c \, \sum_{i,j\in g_s} \frac{q(\vec{x}_i(t)+ \vec{x}_j(t)) - q(\vec{x}_i (t)) - q(\vec{x}_j (t))}{2} 
 = \sum_{s=1}^c q \left(  \sum_{i\in g_s} \vec{x}_i (t) \right),
\end{equation}
which is the equivalent geometric problem given by~\eqref{eq:VP} in a pseudo-Euclidean space.
Therefore the maximisation of the linearised Markov stability~\eqref{eq:linearms}, which is equivalent to the minimisation of the Potts Hamiltonian, can be exactly recovered by a vector partitioning problem on the total magnitude of the sum vectors in a pseudo-Euclidean space. 

As stated above, modularity optimisation is a particular case of this linearisation and corresponds to the maximisation of  $r_{lin}(1,g)$. 

\section{A Vector Partitioning Algorithm Inspired by Graph-theoretical Heuristics}
\label{sec:3}
The partitioning of $n$ vectors in $\mathbb{R}^{n-1}$ into $k$ groups can be solved in time O$(n^{(n-1)(k-1)-1})$~\cite{onn2001vector}. Therefore this problem becomes be infeasible for large networks and approximation methods are needed. Several algorithms have been proposed to address the problem of vector partitioning, including the greedy algorithm~\cite{alpert1995spectral}, fine-tuning following coarse division~\cite{wang2008vector}, and the recently introduced $k$-means-like heuristic~\cite{zhang2015multiway}. All these methods are designed to find an optimised partition given the desired number of groups $s\,\,(2\leq s\leq n)$. To obtain the optimal number of groups, the algorithms need to be used repeatedly to search exhaustively across different values of $s$.

Exploiting the connections between graph partitioning and vector partitioning described above, we present here a different vector partitioning algorithm which gives the optimal number of groups and the associated optimised partition as the output of the algorithm. Our vector partition algorithm is based on the graph-theoretical node agglomeration heuristics used in the popular Louvain algorithm for community detection in networks.


\begin{algorithm}[h!]                    
\caption{Vector Partitioning with the Louvain Heuristic}       
\label{alg:CG}                         
\begin{algorithmic}[1]                 
    \STATE \textbf{Input}: A set of $n$ vertex vectors $\vec{x}_i$
    \STATE Assign every vector to its own group, the group sum vectors $ \vec{y}_i = \vec{x}_i$
    \WHILE{$\mathrm{improvement\_possible}=\mathrm{True}$}
    		\STATE Set $\mathrm{improvement\_possible}=\mathrm{False}$
    		\FOR{$i=1:n$}
    			\STATE Vector $\vec{x}_i$ is currently in group $\alpha$
    			\STATE Compute the change $\Delta r_{\alpha \beta}$ of moving $\vec{x}_i$ from group $\alpha$ to group $\beta$ 
    			\STATE $r_{\mathrm{max}} = \mathrm{max} \, \Delta r_{\alpha \beta}$
    			\STATE $\beta_{\mathrm{max}} = \mathrm{argmax} \, \Delta r_{\alpha \beta}$
    		
    			\IF{$r_{\mathrm{max}}>0$}
        			\STATE $\mathrm{improvement\_possible}=\mathrm{True}$
  				\STATE move $\vec{x}_i$ to group $\beta_{\mathrm{max}}$
  				\STATE $\vec{y}_\alpha = \vec{y}_\alpha - \vec{x}_i$ 
  				\STATE $\vec{y}_{\beta_{\mathrm{max}}} = \vec{y}_{\beta_{\mathrm{max}}} +\vec{x}_i$
    			\ENDIF
    		\ENDFOR
    \ENDWHILE
    \IF{every vector is still partitioned into its own group}
    \STATE Output the partition and return
    \ELSE
        \STATE Use the group sum vectors $\vec{y}_\beta$ as inputs and go to step 2
    \ENDIF
\end{algorithmic}
\end{algorithm}

Community detection is generally defined as an NP-hard optimisation, and several algorithms based on different heuristics have been proposed in recent years~\cite{fortunato2010community}. Among them, the so-called Louvain method~\cite{blondel2008fast} has gained wide use for the optimisation of modularity, Markov Stability and other objective functions, due to its efficiency and good performance against benchmarks. The Louvain method consists of two phases. In the first phase, each vertex is assigned to a different community and each vertex is repeatedly and sequentially moved to its neighbouring community if the gain in the cost function is maximum and positive. The first phase ends if there is no further improvement possible. The second phase builds a new network from the communities found in the first phase, where the nodes of the new network are `supernodes' corresponding to the communities found in the preceding phase. The weight of the links between two supernodes is the sum of the weights between all the nodes in the communities, and the sum of the weights within each community is represented by self-loops. These two phases are repeated iteratively until the network of supernodes is not changed. 

In contrast to purely geometric algorithms for vector partitioning existent in the literature, we adopt a heuristic inspired by the graph-theoretical Louvain method which does not need the declaration of the number of groups \textit{a priori}. The first phase of the algorithm starts with each node vector in its own group, and we compute the gain of the total sum~\eqref{eq:VP} if we move the vector $\vec{x}_i$ from its own group $\alpha$ into another group $\beta$. The vector $\vec{x}_i$ is moved to the group which induces the maximum positive gain. Similarly to the Louvain method, each vector is considered sequentially and repeatedly until there is no possible improvement by a single movement, which means a local optimum is reached. Due to the simplicity of the squared length function, the difference of moving the vector $\vec{x}_i$ from a group $\alpha$ into another group $\beta$ is easily computed as 
\begin{equation}
\label{eq:dr}
\Delta r_{\alpha \beta} = \vec{y}_\beta^T \vec{x}_i  - (\vec{y}_\alpha- \vec{x}_i)^T \vec{x}_i,
\end{equation}
where $\vec{y}_\alpha$ and $\vec{y}_\beta$ are the sum of vectors in groups $\alpha$ and $\beta$, respectively. 
These sum vectors resemble the `supernodes' associated with the communities in the Louvain method,
and are directly used as the input for the next phase of the iteration. The algorithm stops when there are no more changes to the sum vectors. The pseudo-code for the algorithm is shown in Algorithm~\ref{alg:CG}.

The Louvain heuristic leads to a distinct vector partitioning algorithm which has a graph-theoretical origin and inherits the advantages of the Louvain heuristic, i.e., its speed, flexibility, and good performance in practice. This algorithm also applies directly to the vector partitioning problem corresponding to the optimisation of the linearised version of Markov Stability (and hence of the Potts model and modularity). Algorithmically, our vector partitioning implementation has two main differences from the original Louvain method for community detection: (i) the gain in the total sum of one movement can be easily computed as the difference of two inner products; (ii) the sum of the vectors is directly used as the input for the next iteration.

\section{Applications of the Vector Partitioning Algorithm}

To illustrate its use, we apply our vector partitioning algorithm to community detection in an ensemble of random graphs obtained with the Lancichinetti-Fortunato-Radicchi (LFR) benchmark~\cite{lancichinetti2008benchmark}. The LFR graphs all have $n=1000$ nodes with an average degree of 15 and a maximum degree of 50. The minimum and maximum community sizes are 20 and 50, respectively, and the exponents for the degree and community size distributions are 2 and 3, respectively. 
Running the LFR benchmark with these parameters, we obtain 134 graphs that have $k=30$ communities. These graphs constitute our test ensemble used to generate statistically comparable results. 

%
%

\subsection{Approximation of Markov Stability optimisation through vector embeddings of reduced dimensionality}
\label{sec:3.2}

The autocovariance matrix $B(t)$ given in equation.~\eqref{eq:autocov} is the central object in the optimisation of Markov Stability. Its rewriting as a Gram matrix in equation.~\eqref{eq:autocov_Gram} makes it clear that $B(t)$ is a geometric object containing the distances between nodes, as measured in the $(n-1)$-dimensional spectral vector embedding. Equation~\eqref{eq:VP} shows that the optimal graph partition according to Markov Stability is recovered \textit{exactly} by solving the vector partitioning problem when \textit{all} $(n-1)$ non-trivial eigenvectors $\vec{v}_i$ of the matrix $M$ are used to embed the node vectors $\vec{x}_i(t)$.

However, the `distance' (Gram) matrix $B(t)$ can be obtained \textit{approximately} from a lower dimensional embedding involving only a \textit{subset} of the eigenvectors of $M$. This is equivalent to neglecting the components of the node vector associated with small eigenvalues, and solving the vector partitioning problem in a lower dimensional space. 
%
%
%
Such approximations can be crucial numerically when the size of the network is large, such that it is impractical to compute all eigenvectors of $M$. A common approximation in spectral partitioning methods is to use a few eigenvectors corresponding to the leading eigenvalues, and neglect the components associated with small eigenvalues.  As shown in~\cite{newman2006finding}, in order to divide the network into $k$ communities, one must use at least $k-1$ eigenvectors. Using more than $k-1$ eigenvectors will always give more accurate approximations,
yet the improvement achieved by using additional eigenvectors is unclear.
%

\begin{figure}[h!]
\centering
\includegraphics[width= 0.7\linewidth]{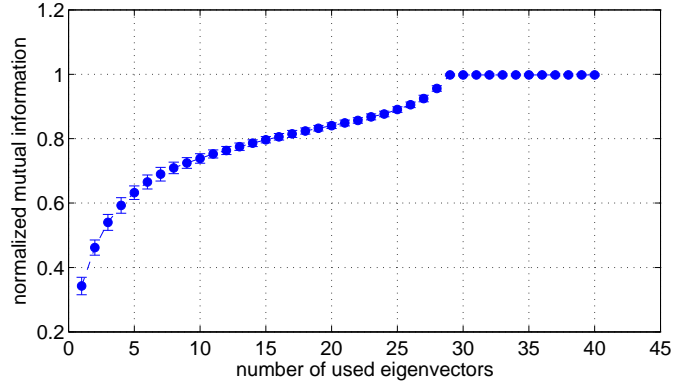}
\caption{The normalised mutual information of the partition obtained from our vector partitioning algorithm compared against the ground truth of the benchmark as a function of the number of eigenvectors used to embed the node vectors. The ground truth partition has $k=30$ communities and the quality of the detected partition does not increase noticeably when using more than $(k-1)=29$ eigenvectors. When the normalised mutual information is 1, the partition obtained is equal to the ground truth. Each point is an average over 134 realizations of the LFR benchmark.}
\label{fig:nmi}
\end{figure}

We have explored this issue numerically by computing the optimised Markov Stability partition for the each of the graphs in the LFR test ensemble, all of which have $k=30$ communities. 
Figure~\ref{fig:nmi} shows the improvement of the quality of the partition obtained with our algorithm (measured as the normalised mutual information~\cite{strehl2002cluster} with respect to the ground truth) as we increase the number of eigenvectors used, i.e., as we increase the dimensionality of the embedding space.
%
%
%
Our numerical experiments show that our vector partitioning for the LFR networks with an underlying $k=30$ community structure, using $k-1$ eigenvectors already gives a highly accurate result and using more eigenvectors does not essentially improve the performance.

\subsection{Spectral maximisation of modularity using the vector partitioning algorithm}
\label{sec:5}
Our results in Section~\ref{sec:4} show that modularity can be maximised using our vector partitioning algorithm applied to a spectral vector embedding based on the eigenvectors of the transition matrix $M = D^{-1}A$ in a pseudo-Euclidean space. 
Interestingly, this embedding is distinct from the traditional spectral methods for modularity maximisation~\cite{newman2006finding,newman2006modularity,zhang2015multiway}, which use instead the eigenvectors of the modularity matrix $B_Q = A - \vec{d}\vec{d}^T/2m$. 
In general, there are several advantages of our formulation with the transition matrix $M$. First, the fact that spectrum of the transition matrix is bounded in $\left[ -1,1\right]$ makes it more suitable for large networks~\cite{bolla2011penalized}. 
Further, if the network is large and sparse, the transition matrix also has sparsity and the decomposition of the matrix can be achieved with fast algorithms, while the modularity matrix is not sparse in general.

\begin{figure}[h!]
\centering
\includegraphics[width= 0.7\linewidth]{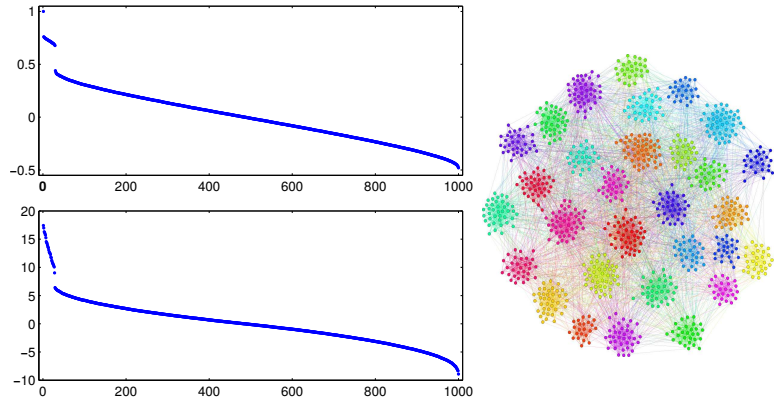}
\caption{Eigenvalues of the transition matrix $M$ (top) and the modularity matrix $B_Q$ (bottom) of one network realisation of the LFR benchmark with 30 communities (right).}
\label{fig:eigens}
\end{figure}

To compare these two spectral methods, we apply our vector partitioning algorithm to maximise modularity on the LFR ensemble using both spectral embeddings: the one based on the eigenvectors of the modularity matrix, and our embedding based on the eigenvectors of the transition matrix. Figure ~\ref{fig:eigens} shows the eigenvalues of $M$ and $B_Q$ for one of the LFR networks with $k=30$ communities. Although both matrices have a spectral gap at $(k-1)$ eigenvectors (the number needed to find the communities), the spectral gap is larger for the transition matrix embedding.

\begin{figure}[hb!]
\centering
\includegraphics[width= 0.7\linewidth]{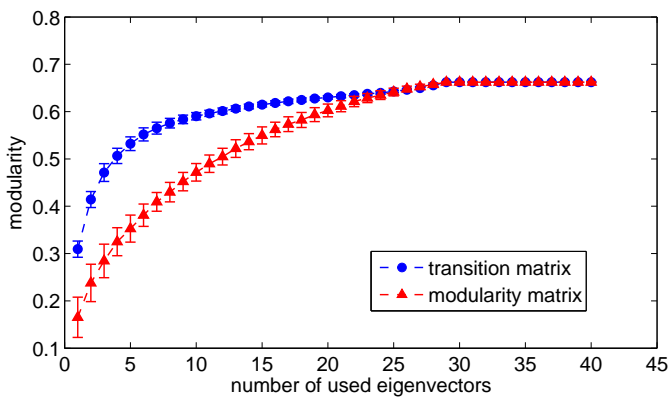}
\caption{Modularity of the optimised partitions obtained with the vector partitioning algorithm using different number of eigenvectors of the transition matrix and the modularity matrix to embed the node vectors. The results are averaged over 134 network realisations of the LFR benchmark, all with $k=30$ communities. As the dimensionality of the embedding increases (i.e., a higher number of eigenvectors), the modularity of the optimised partition increases until it reaches its maximum when $k-1$ eigenvectors are used. Note the higher modularity attained by the spectral embedding based on the transition matrix, as compared to that based on the modularity matrix.}
\label{fig:modularity}
\end{figure}

This cleaner spectral distribution of the transition matrix suggests that modularity maximisation performed on spectral embeddings of reduced dimension (i.e., with just a subset of the eigenvectors) should be more accurate when using the transition matrix eigenvectors than those of the modularity matrix. This is shown in Figure~\ref{fig:modularity}, where we maximise modularity based on embeddings of increasing dimensionality (from 1 to 40) for both sets of eigenvectors.
%
Since all the LFR graphs in the ensemble have 30 communities, the optimum of modularity is reached when using $k-1 = 29$ eigenvectors.
%
%
However, for the same number of eigenvectors (smaller than 29),  the transition matrix embedding provides a higher modularity than the embedding based on the eigenvectors of the modularity matrix itself. 

\begin{figure}[h!]
\centering
\includegraphics[width=0.7\textwidth]{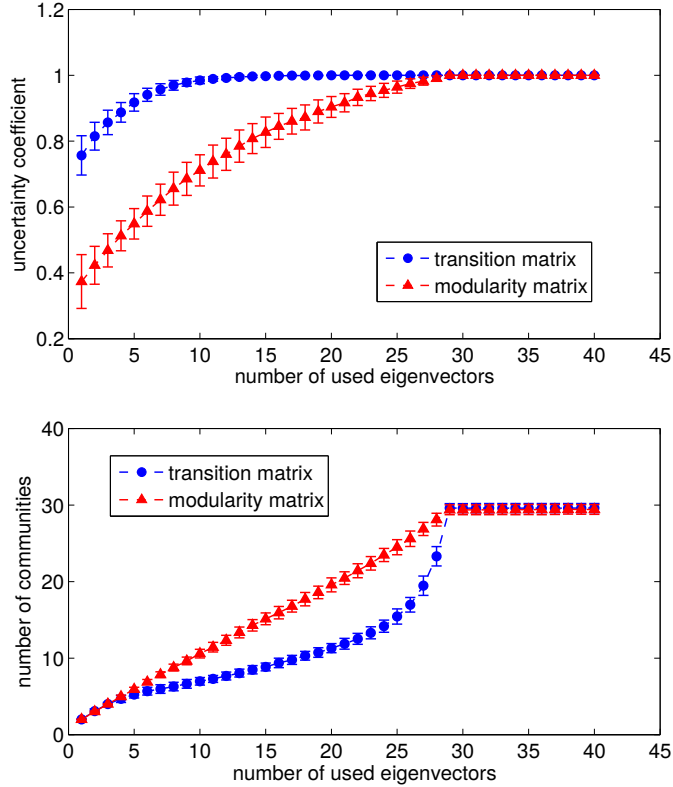} 
\caption{The uncertainty coefficient with respect to the ground truth (top) and the number of communities (bottom) for the optimised partitions obtained with the vector partitioning algorithm using different number of eigenvectors of the transition matrix and the modularity matrix for the spectral embedding of the node vectors. The results are averaged over 134 network realisations of the LFR benchmark, and all the networks have a ground truth of $k=30$ communities.
An uncertainty coefficient of one indicates that 
the detected communities can be obtained by joining communities in the ground truth, i.e., 
there is a hierarchical relationship between the computed partition and the ground truth.
The embedding using the eigenvectors of the modularity matrix produces higher number of communities with lower uncertainty coefficient (with respect to the ground truth).
}
\label{fig:uncertainty}
\end{figure}

Interestingly, the number of communities of the optimised partitions obtained from the two spectral embeddings are substantially different. As shown in Figure~\ref{fig:uncertainty}, the transition matrix spectral embedding gives partitions with fewer communities than the modularity matrix spectral embedding, yet the communities obtained have a higher information content with respect to the ground truth, as indicated by higher uncertainty coefficients~\cite{press2007numerical}. The uncertainty coefficient measures the ratio of the useful information about the ground truth to the total information contained in the computed partition: an uncertainty coefficient of one means that there is a hierarchical relationship between the computed partition and the ground truth.
%
%
%

\begin{figure}[h!]
\centering
\includegraphics[width= 0.7\linewidth]{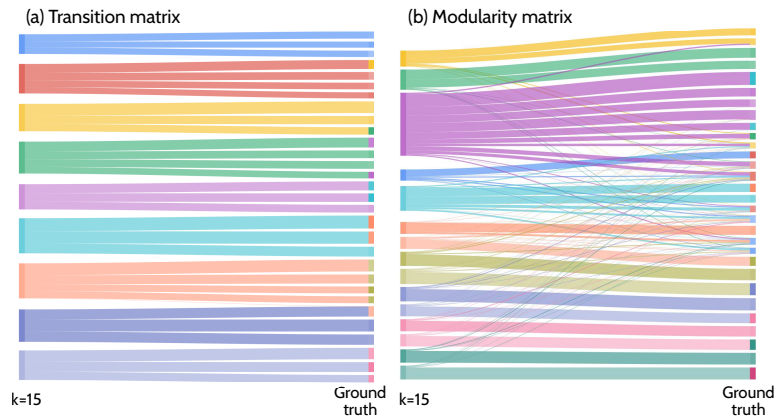}
\caption{A Sankey diagram to show the relationship between the ground truth partition and the partitions obtained with vector partitioning based on a spectral embedding with $k=15$ eigenvectors of (a) the transition matrix and (b) the modularity matrix. The ground truth has 30 communities. With the transition matrix, we get 9 communities which are hierarchical with respect to the ground truth, while with the modularity matrix, we get 15 communities but they do not have a clear hierarchical relationship with the ground truth.}
\label{fig:sankey}
\end{figure}

For example, using 15 eigenvectors for both embeddings, the vector partitioning algorithm obtains a partition into 9 communities for the transition matrix embedding, and into 15 communities for the modularity matrix. However, the uncertainty coefficient of the partition from the transition matrix embedding is substantially higher. 
As shown explicitly in the Sankey diagram in Figure~\ref{fig:sankey}, the partition obtained with the transition matrix embedding is coarser, yet fully compatible with the ground truth, whereas the finer partition obtained from the modularity matrix embedding presents marked inconsistencies with the ground truth partition.  
This indicates that, compared to the modularity matrix, more information about the community structure is contained in the leading eigenvectors of the transition matrix. This property has more significance when the network is large and only a few leading eigenvalues and eigenvectors can be computed. With the transition matrix, we can get a more informative partition of the network using fewer eigenvectors. 

\section{Conclusion}
\label{sec:6}

In this paper, we have presented a connection between vector partitioning, spectral embeddings and Markov Stability, a generalised framework for multi-scale community detection based on the dynamics of Markov processes on graphs. We show that Markov Stability community detection in a graph with $n$ nodes corresponds to a vector partitioning problem in 
$\mathbb{R}^{n-1}$, where the nodes of the graph are represented by time-dependent vectors in an $(n-1)$-dimensional embedding spanned by the eigenvectors of the transition matrix associated with the graph.  
The time of the Markov process plays the role of an inhomogeneous geometric resolution factor, which acts differently along the different components of the node vectors.
This feature allows for the exploration of changes in optimal groupings at different resolutions leading to potential multi-scale community detection. We also show that the Reichardt \& Bornholdt Potts model and modularity optimisations for community detection correspond to a vector partitioning problem in a pseudo-Euclidean space. 
In addition, the vector representation provides a clear explanation of the distinct quantity optimised by $k$-means and establishes the connection to the usual spectral clustering.

Exploiting the graph-theoretical interpretation, we then propose a vector partitioning algorithm using a heuristic inspired by the Louvain method for community detection in networks. Application of our algorithm on benchmark networks shows that for a network with $k$ communities, vector partitioning with only $k-1$ eigenvectors is enough to reveal the community structure. Finally, we compare spectral methods for community detection based on the embedding provided by the eigenvectors of the transition matrix and the modularity matrix. In the sense of modularity, the information about the community structure is more compressed in the leading eigenvectors of the transition matrix. Thus vector partitioning with the transition matrix is capable of unveiling the community structure with fewer eigenvectors.


A number of questions would be of interest for further investigation.
Our decomposition of the Markov Stability matrix $B(t)$ as a Gram matrix is key to the link with geometrical interpretations and opens the possibility of using other kernels from the graph~\cite{scholkopf2001learning}. 
Indeed, it would be worth exploring how the distance associated with the Markov Stability kernel is related to other graph kernels, such as the kernel of the pseudoinverse of the Laplacian matrix which preserves the average commute time distance and the diffusion map which preserves the diffusion distance~\cite{fouss2007random,coifman2006diffusion}. 
Mapping the nodes onto a vector space could also be used to find soft partitions for overlapping communities as a projection problem rather than a vector partition problem, thus recasting this problem geometrically.  We leave these interesting problems as open directions for future work.

\section*{Funding}
This work was supported by the European Commission [European Union 7th Framework Programme for research, technological development and demonstration under grant agreement no. 607466]; and the Engineering and Physical Sciences Research Council [EP/N014529/1 to M.B.].

\section*{Acknowledgement}
The authors would like to thank Dr Michael Schaub for extended discussions, and Dr Sam Greenbury for reading the manuscript and giving useful comments. We also thank the two reviewers for helpful suggestions.

\bibliographystyle{ieeetr}
\bibliography{references}
%


\end{document}